\documentclass[preprint,aps,prl,superscriptaddress]{revtex4-2}

\usepackage[pdftex]{graphicx}
\usepackage{amsmath}

\begin{document}

\title{Quantum decoherence by magnetic fluctuations in a magnetic topological insulator}

\author{Ruben Saatjian}
\affiliation{Materials Department, University of California, Santa Barbara, California 93106, USA}

\author{Simon Dovr\'{e}n}
\affiliation{Materials Department, University of California, Santa Barbara, California 93106, USA}

\author{Kohtaro Yamakawa}
\affiliation{Department of Physics, University of California, Berkeley, California 94720, USA}

\author{Ryan S. Russell}
\affiliation{Materials Department, University of California, Santa Barbara, California 93106, USA}

\author{James G. Analytis}
\affiliation{Department of Physics, University of California, Berkeley, California 94720, USA}

\author{John W. Harter}
\email[Corresponding author: ]{harter@ucsb.edu}
\affiliation{Materials Department, University of California, Santa Barbara, California 93106, USA}

\date{\today}

\begin{abstract}
In magnetic topological insulators, spontaneous time-reversal symmetry breaking by intrinsic magnetic order can gap the topological surface spectrum, resulting in exotic properties like axion electrodynamics, the quantum anomalous Hall effect, and other topological magnetoelectric responses. Understanding the magnetic order and its coupling to topological states is essential to harness these properties. Here, we leverage near-resonant magnetic dipole optical second harmonic generation to probe magnetic fluctuations in the candidate axion insulator EuSn$_2$(As,P)$_2$ across its antiferromagnetic phase boundary. We observe a pronounced dimensional crossover in the quantum decoherence induced by magnetic fluctuations, whereby two-dimensional in-plane ferromagnetic correlations at high temperatures give way to three-dimensional long-range order at the N\'eel temperature. We also observe the breaking of rotational symmetry within the long-range-ordered antiferromagnetic state and map out the resulting spatial domain structure. More generally, we demonstrate the unique capabilities of nonlinear optical spectroscopy to study quantum coherence and fluctuations in magnetic quantum materials.
\end{abstract}

\maketitle

\section{Introduction}

Magnetic topological insulators are a class of quantum materials that combine nontrivial electronic topology with magnetic order~\cite{tokura2019}. Spontaneous time-reversal symmetry breaking by magnetism, or by magnetic fluctuations sufficiently slower than the characteristic electronic timescale, can promote a range of novel phenomena, including axion electrodynamics~\cite{essin2009,li2010,mogi2017,xu2019,zhang2019,sekine2021}, the quantum anomalous Hall effect~\cite{mong2010,liu2008,yu2010,chang2013}, chiral topological superconductivity~\cite{qi2010}, and the generation of ideal Weyl nodes~\cite{ma2019,soh2019,jo2020}. These states have unique properties with potential applications in spintronics and other advanced technologies. The first experimental observation of such physics occurred in conventional topological insulators extrinsically doped with dilute magnetic ions~\cite{liu2008,yu2010,chang2013}, but this necessarily introduces substantial disorder into the material. More recently, intrinsic magnetic topological insulators, which are stoichiometric compounds with native long-range magnetic order, have garnered interest as clean and tunable material platforms that may facilitate the observation and study of topological magnetoelectric phenomena at higher temperature scales.

\begin{figure*}[t]
\includegraphics[width=6.5in]{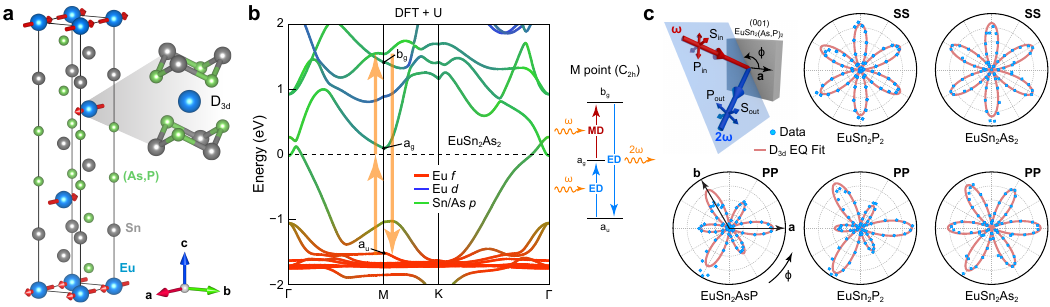}
\caption{\label{figure1} Crystal and electronic structure of EuSn$_2$(As,P)$_2$ and RA-SHG patterns. \textbf{a}~Crystal structure showing half of the conventional magnetic unit cell. Red arrows depict Eu$^{2+}$ local magnetic moments, which lie within the plane along the $a$ axis. Moments are aligned within each plane and antialigned with neighboring planes, forming long-range A-type antiferromagnetism. The inset illustrates the local environment surrounding the Eu$^{2+}$ ions, showing $D_{3d}$ site symmetry. \textbf{b}~Orbital-resolved electronic band structure of EuSn$_2$As$_2$ from DFT. Vertical arrows highlight optical transitions near the $M$ point contributing to near-resonant SHG. The irreducible representations of the states participating in the resonances are labeled. The right inset shows an energy level diagram of MD-SHG involving both ED and MD optical transitions. \textbf{c}~Parallel polarization RA-SHG patterns for all three materials at 50~K. The top left panel defines the oblique incidence scattering geometry. The top row shows S$_\mathrm{in}$-S$_\mathrm{out}$ (SS) polarization patterns and the bottom row shows P$_\mathrm{in}$-P$_\mathrm{out}$ (PP) polarization patterns.}
\end{figure*}

A series of Eu-based intrinsic magnetic topological insulators, including EuCd$_2$As$_2$, EuSn$_2$As$_2$, and EuSn$_2$P$_2$, has emerged in recent years~\cite{ma2019,soh2019,jo2020,arguilla2017,rahn2018,gui2019,li2019b,pierantozzi2022}. This family has a layered structure with Eu$^{2+}$ ions forming a triangular lattice sandwiched between two covalently-bonded buckled honeycomb layers, such as (SnAs)$_2^{2-}$, from which the topological bands are derived. Magnetism arises from half-filled $4f^7$ subshells on the Eu$^{2+}$ ions. These localized high-spin ($S = 7/2$) moments align ferromagnetically in the plane and antiferromagnetically out of the plane (Fig.~\ref{figure1}a), resulting in long-range A-type antiferromagnetism below a N\'eel temperature $T_N$ that ranges from 10~K to 30~K. The distinct spatial separation of magnetic and topological electronic states in these materials raises the possibility of precise magnetic control over the electronic topology, for example through the direction of magnetization~\cite{xu2019}.

In this work, we study fluctuations and symmetry breaking by the magnetic degrees of freedom in the Eu-based materials using near-resonant magnetic dipole optical second harmonic generation (SHG). We focus on the compounds EuSn$_2$As$_2$, EuSn$_2$P$_2$, and EuSn$_2$AsP, referred to collectively as EuSn$_2$(As,P)$_2$, which have the highest reported $T_N$ (24~K and 30~K for the As and P endmembers, respectively~\cite{arguilla2017,gui2019}). We measure the quantum decoherence induced by magnetic fluctuations across $T_N$ and find a dimensional crossover from short-range two-dimensional in-plane ferromagnetic correlations above $T_N$ to three-dimensional long-range antiferromagnetic order below it. Within the low-temperature antiferromagnetic phase, we detect broken rotational symmetry due to global alignment of the in-plane magnetic moments, and map out the corresponding spatial domain patterns. Our results shed light on the magnetism in this series of materials, paving the way for the potential control of topological magnetoelectric properties. More broadly, we showcase the capabilities of nonlinear optical spectroscopy to access the degree of quantum coherence of magnetic fluctuations in quantum materials.

\section{Results}

Single crystals of EuSn$_2$As$_2$, EuSn$_2$P$_2$, and the 1:1 alloy EuSn$_2$AsP were prepared by a flux method similar to Refs.~\citenum{pakhira2021} and \citenum{riberolles2021}. All three materials show similar nonlinear optical responses, and magnetic susceptibility measurements show clear signatures of antiferromagnetic transitions (Supplementary Note~1), with $T_N$ closely matching previously reported values~\cite{arguilla2017,gui2019,li2019b,pakhira2021}. Despite the similar intrinsic behavior of the three compounds, EuSn$_2$As$_2$ yielded the largest single crystals, and we focus on this material when mapping spatial domain patterns. Measuring the rotational anisotropy of optical SHG (RA-SHG) is a precise method to access the nonlinear optical susceptibility tensor and deduce the point group symmetries of a material. Using RA-SHG, tensor elements can be tracked across a wide temperature range to reveal minute changes in order, symmetry, dimensionality, and the strength of interactions in a sample~\cite{harter2017,orenstein2021,ron2019,ahn2024}. The extreme sensitivity of SHG can be further enhanced near resonances, strengthening the SHG response by orders of magnitude and amplifying signatures of states involved in the resonant transition. As Fig.~\ref{figure1}b demonstrates, 800~nm excitation (1.55~eV) is nearly resonant with several energy transitions involving Eu $4f$ and $5d$ orbitals at the $M$ point, gaining us direct access to the magnetic degrees of freedom.

Figure~\ref{figure1}c illustrates the RA-SHG experimental scattering geometry and corresponding data for all three samples. In the S$_\mathrm{in}$-S$_\mathrm{out}$ channel, six equal lobes appear, while in the P$_\mathrm{in}$-P$_\mathrm{out}$ channel, the underlying 3-fold rotational symmetry of the crystallographic point group $D_{3d}$ ($\bar{3}m$) is revealed through alternating large and small lobes. The measured SHG intensity is relatively strong, which is unusual because $D_{3d}$ is centrosymmetric and leading-order bulk electric dipole (ED) SHG---the usual source of strong nonlinear optical responses---is hence excluded by symmetry. Among the most likely higher-order nonlinear responses are surface ED, bulk magnetic dipole (MD), and bulk electric quadrupole (EQ) SHG~\cite{orenstein2021}. Surface SHG typically arises from slight atomic distortions at the surface of a material and is therefore generally weakly dependent on temperature. We exclude this possibility based on the striking temperature dependence that we observe and discuss below. Thus, bulk MD-SHG or EQ-SHG is the primary nonlinear response channel in EuSn$_2$(As,P)$_2$.

MD-SHG and EQ-SHG are both described by a fourth-rank nonlinear optical susceptibility tensor $\chi_{ijkl}$ that relates the induced nonlinear polarization density in a material $P_i(2\omega)$ to two powers of the incident electric field $E_i(\omega)$ via the equation $P_i(2\omega) = \chi_{ijkl} E_j(\omega) \nabla_k E_l(\omega)$~\cite{orenstein2021}. Within point group $D_{3d}$ and with additional permutation symmetries adapted from Ref.~\citenum{yang2009}, we may calculate the explicit angular dependence of the RA-SHG patterns. For the two parallel polarization channels shown in Fig.~\ref{figure1}c, we obtain $I_{SS}(\phi) = |\alpha\sin(3\phi)|^2$ and $I_{PP}(\phi) = |\alpha\cos(3\phi) + \beta|^2$, where $\alpha = \chi_{xxzx}\cos(\theta)$, $\beta = (\chi_{xxxx} - \chi_{xxzz})\sin(\theta)$, $\phi$ is the azimuthal angle of the scattering plane, and $\theta$ is the angle of incidence relative to the sample surface normal (Supplementary Note~2). Fits of the data to these functional forms show excellent agreement, as demonstrated in Fig.~\ref{figure1}c.

\begin{figure}[t]
\includegraphics{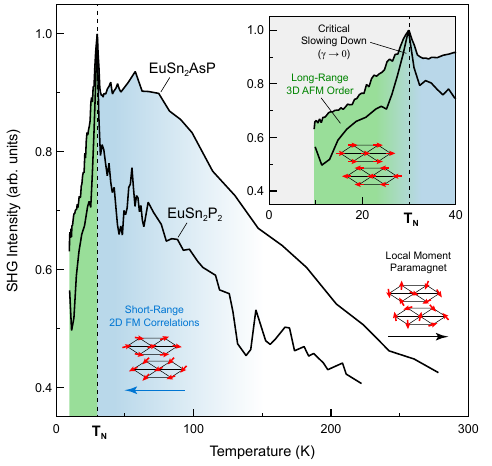}
\caption{\label{figure2} Temperature dependence of SHG. Upon cooling from room temperature, the SHG intensity increases dramatically. Below $\sim$100~K, a broad hump (most apparent in EuSn$_2$AsP) and large fluctuations (most apparent in EuSn$_2$P$_2$) are observed. We identify these features with the emergence of fluctuating short-range ferromagnetic correlations within the Eu planes (blue). Within $\sim$2~K of $T_N$, a pronounced peak appears, which we attribute to critical slowing down at the magnetic phase transition. Below $T_N$, the SHG intensity decreases as long-range three-dimensional antiferromagnetic order freezes in (green). The inset shows a close-up view of the low-temperature region.}
\end{figure}

A defining characteristic of MD-SHG and EQ-SHG is an optical transition between two states of the same parity, a good quantum number in centrosymmetric systems. Such excitations are forbidden for conventional ED transitions. With this symmetry condition in mind, we examine the irreducible representations of wavefunctions calculated by density functional theory (DFT) for EuSn$_2$As$_2$ and identify the $M$ point as the likely source of near-resonant SHG at 800~nm. At this high-symmetry point, with little group $C_{2h}$ ($2/m$), we find both MD and EQ transitions are allowed, but the amplitude of one particular pathway involving a MD transition exceeds all others by an order of magnitude (Supplementary Note~3). Thus, we conclude that MD-SHG is the primary nonlinear response channel in EuSn$_2$(As,P)$_2$. An electron occupying a Eu $f_{xyz}$, $f_{xz^2}$, or $f_{x(x^2-3y^2)}$ orbital (odd-parity $a_u$ symmetry) can be excited to an unoccupied even-parity $a_g$ virtual state through resonant ED absorption at $\sim$1.5~eV, further excited by a parity-preserving MD transition to a higher $b_g$ virtual state, and then return to its initial state through resonant ED emission at $\sim$3~eV. While other $k$-points will contribute weakly to the nonlinear optical response, none possess the same combination of orbital symmetry and double resonance necessary for strong SHG. The overall resonant MD-SHG process is illustrated in Fig.~\ref{figure1}b. The participating unoccupied virtual states have partial ($\sim$20\%) Eu $5d$ character but majority Sn and As $p$ character. Thus, in real space we may interpret the initial optical transition as a charge-transfer excitation in which an electron is removed from a tightly-bound Eu $4f$ orbital and added to the extended (SnAs)$_2$ network. This leaves behind a Eu atom in a $4f^6$ configuration with a reduced spin. The reduced moment on the spin lattice represents a localized magnetic excitation that will propagate outward as it becomes entangled with neighboring spin. This process of magnetic decoherence of the virtual states is expected to be sensitive to the strength and nature of the magnetic order in the system, as we detail below.

Now that near-resonant MD-SHG has been established as the dominant nonlinear optical response, we turn to its temperature dependence. Figure~\ref{figure2} shows SHG intensity integrated over $\phi$ versus temperature for EuSn$_2$P$_2$ and EuSn$_2$AsP, which share a similar $T_N$. Data for EuSn$_2$As$_2$ is shown in Supplementary Note 7. Several remarkable features are apparent in the data for all three compounds. First, we observe a significant growth in SHG as the sample is cooled, more than doubling between room temperature and 100~K. Second, below $\sim$100~K a broad high-intensity hump exists with large associated fluctuations. Third, upon traversing the magnetic transition temperature $T_N$, an extremely sharp SHG peak occurs. Finally, as temperature is further lowered in the magnetically-ordered phase, the SHG intensity decreases again. Such behavior resembles a linear magnetic susceptibility, but we emphasize that we are here measuring a nonlinear optical susceptibility not directly sensitive to the magnetic degrees of freedom. Generally, one may expect a slight change in the SHG intensity below the magnetic ordering temperature, but that is not what we observe. Instead, we appear to be sensitive to \textit{fluctuations} of the magnetic order. In particular, we observe an increase in SHG likely associated with the growth of short-range two-dimensional ferromagnetic correlations within the individual Eu layers above $T_N$, giving rise to a broad hump with large accompanying fluctuations. In EuSn$_2$(As,P)$_2$ and related materials, the in-plane ferromagnetic exchange is much stronger than the out-of-plane antiferromagnetic exchange because of the distances between Eu atoms, and it is therefore natural to expect in-plane correlations to emerge before out-of-plane order condenses~\cite{ma2019,sun2021,blawat2022}. Large fluctuations are expected in this regime, as neighboring planes are only weakly coupled. Indeed, in the limit of completely isolated planes, true long-range two-dimensional order is suppressed by fluctuations~\cite{mermin1966}. Strong in-plane ferromagnetic fluctuations below $\sim$100~K have also been detected in the related material EuCd$_2$As$_2$ by muon spin relaxation, and it has been suggested that these short-range magnetic fluctuations are sufficient to induce topological properties even in the absence of long-range magnetic order~\cite{ma2019,yi2023}. At $T_N$, we observe a dimensional crossover as the in-plane ferromagnetic correlations give rise to true long-range three-dimensional antiferromagnetic order. Here, critical fluctuations are expected to be largest, and this is reflected in a sharp SHG peak. Finally, upon further cooling, the long-range magnetic order grows in strength and fluctuations accordingly diminish, as does the SHG intensity. Thus, taken as a whole, our observations are consistent with a pronounced sensitivity of MD-SHG to fluctuations of the magnetic degrees of freedom, rather than the magnetic order itself.

To interpret this result, we examine the expression for the MD-SHG nonlinear optical susceptibility tensor derived from perturbation theory~\cite{boyd2020}:
\begin{widetext}
\begin{equation*}
\chi_{ijkl}(2\omega, \omega, \omega) = \frac{N}{2\epsilon_0\hbar^2}\sum_{mvn}\left(\rho^\mathrm{(eq)}_{mm} - \rho^\mathrm{(eq)}_{vv}\right)\frac{\left\langle m\right|\hat\mu_i\left|n\right\rangle\left\langle n\right|\hat{Q}_{kl}\left|v\right\rangle\left\langle v\right|\hat\mu_j\left|m\right\rangle}{\left[(\omega_{nm} - 2\omega) - i\gamma_{nm}\right]\left[(\omega_{vm} - \omega) - i\gamma_{vm}\right]},
\end{equation*}
\end{widetext}
where $N$ is the electron density, $m$, $v$, and $n$ label eigenstates, $\rho^\mathrm{(eq)}_{nm}$ is the equilibrium density matrix, $\hat\mu_i$ and $\hat{Q}_{ij}$ are components of the ED and MD operators, respectively, $\omega_{nm} = (E_n - E_m)/\hbar$, and $\gamma_{nm}$ is a phenomenological parameter introduced to account for environmental degrees of freedom that contribute to quantum decoherence but are not explicitly included in the density matrix. More precisely, the density matrix obeys the quantum master equation
$${\dot{\rho}_{nm} = \frac{1}{i\hbar}\left[\mathcal{H},\rho\right]_{nm} - \gamma_{nm}\left(\rho_{nm} - \rho^\mathrm{(eq)}_{nm}\right),}$$
showing that $\gamma_{nm}$ parameterizes quantum coherence through the rate at which off-diagonal density matrix elements decay in time: $\rho_{nm}(t) \propto e^{-i\omega_{nm}t}e^{-\gamma_{nm}t}$. Here, the density matrix captures orbital (charge) eigenstates but not magnetism, so quantum decoherence primarily arises from interactions with the background magnetic environment. As shown above, the SHG signal is dominated by a near resonance at the $M$ point, where both $\omega_{nm} \approx 2\omega$ and $\omega_{vm} \approx \omega$. This significantly amplifies the effects of the $\gamma$ parameters appearing in the denominator of the susceptibility expression. In fact, these are the only parameters that can appreciably influence the temperature dependence of SHG, as $\rho^\mathrm{(eq)}_{mm} - \rho^\mathrm{(eq)}_{vv} = 1$ because $m$ is occupied and $v$ is unoccupied, and the ED and MD matrix elements in the numerator are not expected to change significantly with temperature. Thus, we see that near-resonant MD-SHG is acutely sensitive to the quantum decoherence induced by magnetic fluctuations in the system, as we have inferred from our temperature dependence measurements.

A hallmark of second-order phase transitions is ``critical slowing down,'' a significant increase in the time it takes for a system to respond to perturbations and return to equilibrium. Within the context of magnetic transitions, this occurs when locally-fluctuating moments become entangled with their neighbors at increasingly long distances. Both the correlation length $\xi \propto \left|T - T_c\right|^{-\nu}$ and time $\tau \propto \xi^z$ diverge at $T_N$, with critical exponents $\nu$ and $z$, respectively~\cite{sethna2006}. In Supplementary Note~4, we derive the relationship between the off-diagonal quantum decoherence rate $\gamma_{nm}$ and the magnetic correlation time: $\gamma^{-1}_{nm} = 2\tau$. This results in $\gamma_{nm} \rightarrow 0$ at $T_N$, which we propose as the likely origin for the sharp peak in SHG that we observe. We emphasize that within this picture, the growth of SHG approaching $T_N$ results from an increase in the \textit{lifetime} of magnetic fluctuations, not their amplitude. We can place a rough upper bound on the coherence time $\tau$ of the $4f^7 \rightarrow 4f^6$ charge-transfer magnetic excitation. We observe an $O(1)$ change in SHG over the measured temperature range, which implies that $\gamma$ must be at least of the same order as $\Delta\omega$, the spectral width of the ultrafast laser. Thus, we calculate $\tau \sim 1/\gamma < 1/\Delta\omega \sim 16$~fs. This value is comparable to the excitation lifetime of an isolated magnetic atom bound to a metallic substrate, as measured by inelastic tunneling spectroscopy, where it was suggested that the short lifetime was due to hybridization with metallic bands, leading to efficient electron-electron interactions that relax the magnetic state of the atom~\cite{schuh2010}. EuSn$_2$(As,P)$_2$ is also metallic, and it has been suggested that the magnetic exchange coupling in the material is mediated by conduction electrons via the Ruderman–Kittel–Kasuya–Yosida (RKKY) interaction~\cite{arguilla2017,bi2022}. Indeed, the electrical conductivity of each compound directly correlates with $T_N$ (Supplementary Note 1), supporting a picture of carrier-mediated exchange. Thus, we may also expect a similarly short magnetic excitation lifetime due to coupling to conduction electrons.

According to the fluctuation-dissipation theorem, there is a close relationship between the imaginary part of a linear susceptibility and the power spectral density of fluctuations of that susceptibility's conjugate observable~\cite{sethna2006}. No such theorem exists for nonlinear susceptibilities, where the imaginary part does not represent dissipation~\cite{pershan1963}. It has been theoretically shown, however, that nonlinear responses are much more sensitive to semiclassical chaotic behavior through the interference of multiple dynamical Liouville-space trajectories~\cite{mukamel1996}. The second-order susceptibility, for example, is directly related to the stability matrix of a system and its associated Lyapunov exponents quantifying the system's sensitivity to initial conditions. Our measurements largely affirm this notion, demonstrating that, at least near resonances, the nonlinear optical response of a material can be significantly influenced by the dynamics of fluctuations of some order parameter of the material, even when that order is not the conjugate observable. We expect that this principle is generally applicable, and that under resonant conditions the nonlinear optical response can be harnessed to study fluctuations, dynamics, and quantum coherence across many types of phase transitions. In the case of magnetic systems, this information is complementary to that of the conventional linear optical susceptibility, whose real and imaginary parts have already been shown to be sensitive to lifetimes and entanglement of spin excitations in exotic magnets~\cite{zhang2024}.

\begin{figure}[t]
\includegraphics{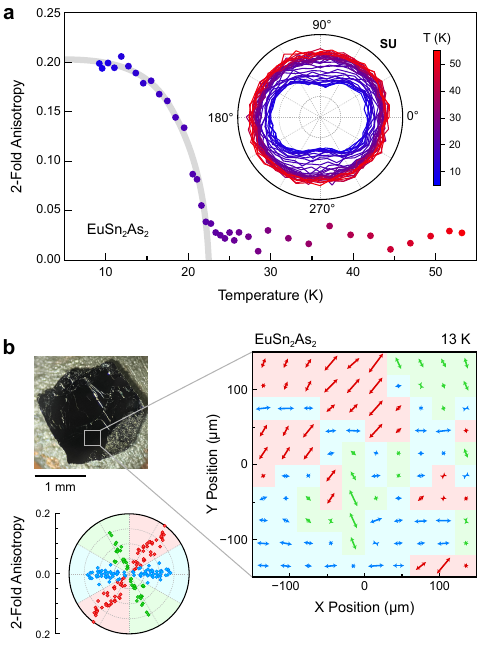}
\caption{\label{figure3} Spontaneous rotational symmetry breaking in EuSn$_2$As$_2$. \textbf{a}~Temperature dependence of the 2-fold rotational anisotropy that emerges at the antiferromagnetic phase transition. The inset shows S$_\mathrm{in}$-U$_\mathrm{out}$ (SU) RA-SHG patterns for each temperature, and the main panel shows the corresponding 2-fold anisotropy versus temperature. The gray curve is a guide to the eye, showing an order-parameter-like onset below $T_N$. \textbf{b}~Spatial map of rotational symmetry breaking. RA-SHG patterns were acquired on a $10\times10$ grid spanning a $300\times300$~$\mu$m area. At each grid point, the amplitude and direction of the 2-fold anisotropy was extracted (represented by the length and direction, respectively, of the double-headed arrows). The upper left inset shows a photograph of the sample with the mapped region identified, and the lower left inset shows a polar plot of the anisotropy sorted by direction into three symmetry-equivalent antiferromagnetic domains (red, green, and blue).}
\end{figure}

Below $T_N$, the Eu$^{2+}$ moments globally align, spontaneously breaking the $\bar{3}$ symmetry axis of the crystal as they choose to orient along one of three equivalent crystallographic directions. Rotational symmetry breaking can be revealed through RA-SHG patterns, which are exceptionally sensitive to point group symmetries~\cite{orenstein2021}. As Fig.~\ref{figure3}a demonstrates, we indeed observe rotational symmetry breaking below $T_N$, with a clear order-parameter-like onset. We measure the unpolarized SHG output (U$_\mathrm{out}$) because a fully isotropic response is expected in the high-symmetry phase, degenerating into a 2-fold ``peanut-like'' shape when the symmetry axis is lost (Supplementary Note~5). We attribute this observation to two possible mechanisms, both involving spin-orbit coupling. In the first, coupling of the globally-aligned spins to the initial-state $f$ orbitals ($f_{xyz}$, $f_{xz^2}$, and $f_{x(x^2-3y^2)}$) shifts their energies, and this manifests as slight changes to the optical resonance condition for different polarization angles, and in the second, global alignment of moments influences the quantum coherence of different initial-state $f$ orbitals. Accompanying spontaneous symmetry breaking is the formation of magnetic domains transforming into each other by the broken symmetry element. Optical imaging of antiferromagnetic domains is typically challenging because of the absence of a net magnetization, with SHG being one of the few known techniques for doing so~\cite{fiebig2005,seyler2022}. Figure~\ref{figure3}b shows a spatial RA-SHG mapping over a $300\times300$~$\mu$m area that reveals these domains. We find a roughly equal distribution of all three possible domain orientations, with a domain length scale of $\sim$100~$\mu$m. Thermal cycling across $T_N$ generates new domain patterns (Supplementary Note~6), suggesting domain boundaries are not strongly pinned by disorder.

\section{Discussion}

In conclusion, we harnessed near-resonant MD-SHG to probe quantum decoherence by magnetic fluctuations in the candidate axion insulator EuSn$_2$(As,P)$_2$. We observed an unusually sharp SHG peak at the magnetic phase boundary that we attribute to critical slowing down, where the dynamics of the system become extremely slow and the correlation length and time diverge. The quantum coherence time of the charge-transfer $4f^7 \rightarrow 4f^6$ magnetic excitation is estimated to be no more than $\sim$16~fs. We conjecture that such a short lifetime is due to efficient coupling between the localized magnetic moments and the conduction electrons, supporting a scenario whereby RKKY interactions mediate magnetic exchange~\cite{arguilla2017,bi2022}. This conclusion has important implications for the coupling strength between the magnetic order and the topological electronic states, where prior work has reported only a very weak hybridization~\cite{li2019b}. We also uncovered evidence of strong short-range in-plane ferromagnetic correlations surviving far above $T_N$, which in similar materials has been proposed to stabilize topological phases~\cite{ma2019} and has also been attributed to the possible formation of magnetic polarons~\cite{zhang2020}. Finally, we revealed the clear breaking of rotational symmetry within the antiferromagnetic phase and used it to map out the spatial domain structure. As a whole, our findings help to clarify the nature of the magnetic order in this class of materials and its interactions with topological surface states, which play a critical role in realizing topological magnetoelectric phenomena. More generally, our work shows that despite a lack of symmetry breaking enabling a direct coupling of the antiferromagnetic order parameter to SHG, resonant SHG can nevertheless still couple to magnetic fluctuations, providing access to antiferromagnetic order through an entirely different mechanism than considered by previous SHG studies of centrosymmetric antiferromagnets~\cite{fiebig2001}. Ultimately, our work showcases the unique capabilities of nonlinear optical spectroscopy as a powerful tool for investigating quantum coherence, fluctuations, and symmetry breaking in magnetic quantum materials.

\section{Methods}

\subsection{Optical experiments}
SHG experiments were performed on freshly-cleaved samples in an optical cryostat using a laser supplying 40~fs pulses at a 50~kHz repetition rate with a center wavelength of 800~nm and a fluence of $\sim$1~mJ/cm$^2$. The oblique-incidence experimental geometry is illustrated in the upper left panel of Fig.~\ref{figure1}c. The incidence angle relative to the sample surface normal was approximately 10$^\circ$. For the RA patterns shown in Fig.~\ref{figure1} and the temperature dependence shown in Fig.~\ref{figure2}, the SHG signal was measured using a scientific CMOS camera. RA patterns were collected by directly rotating the sample, with the scattering plane remaining fixed. For the RA patterns and domain scans shown in Fig.~\ref{figure3}, the SHG signal was measured using a silicon photomultiplier detector and a lock-in amplifier. For these measurements, normal incidence was used and the polarization of the incident beam was rotated instead of the sample, which afforded a much higher spatial resolution for mapping.

\subsection{Density functional theory calculations}
DFT calculations were performed with the Vienna Ab initio Simulation Package (\textsc{vasp}) using the supplied projector augmented wave potentials within the generalized gradient approximation, the experimental structural parameters~\cite{arguilla2017}, a 400~eV energy cutoff, and a $6 \times 6 \times 6$ k-point mesh. Following prior work, to match the experimental energy position of the Eu $4f$ bands, a Hubbard parameter $U = 5.0$~eV was employed~\cite{li2019b}.

\subsection{Sample synthesis and characterization}
High quality single crystals of EuSn$_2$As$_2$ were prepared by a flux method similar to Ref.~\citenum{pakhira2021}. Europium (99.9999\%, Ames Lab) was cut up and combined with arsenic pieces (99.99\%, Alfa Aesar) and Tin shot (99.9999\%, Thermo Scientific) in a 1:3:20 molar ratio with a total mass of 4.0~g in an alumina crucible under a dry nitrogen glove box atmosphere. The crucible was sealed in a quartz ampoule under 300~Torr of argon without being exposed to air. The ampoule was heated in a box furnace from room temperature to 850$^\circ$C over 10 hours, held for 24 hours, then heated to 980$^\circ$C over 24 hours, before soaking at 980$^\circ$C for 12 hours. Finally, the ampoule was slowly cooled to 650$^\circ$C at 3.67$^\circ$C/hour, and spun in a centrifuge to remove excess tin flux. Shiny crystals in the form of hexagonal platelets with a typical diameter of 2~mm were obtained. Crystals of EuSn$_2$P$_2$ and EuSn$_2$PAs were similarly prepared but with a different peak soaking temperature of 1000$^\circ$C and a centrifuging temperature of 700$^\circ$C, yielding crystals with a typical diameter of 1~mm. EuSn$_2$As$_2$ crystals grown without the heat soaks and the argon backfill pressure had notably smaller diameters, at around 200~$\mu$m. Both of these procedures have been used in previous growths of EuIn$_2$As$_2$~\cite{riberolles2021} to help avoid any high vapor pressure of arsenic and increase the homogeneity of the flux solution before reaching the peak temperature. We also note that growth attempts which centrifuged a Eu-Sn-As mixture at 750$^\circ$C yielded Eu$_5$Sn$_2$As$_6$, a black needle-like crystal with a width of 100~$\mu$m and a length of a few mm. The structure and quality of the crystals was verified using powder x-ray diffraction on a Rigaku Ultima-4 system as well as energy-dispersive x-ray spectroscopy (EDX/EDS) on a SCIOS 2 DualBeam system.

\section{Data Availability}

The data that support the findings of this study are available from the corresponding author upon reasonable request.

\section{Acknowledgments}

This work was supported by the U.S. Air Force Office of Scientific Research (AFOSR) under Award No.~FA9550-22-1-0270. The research reported here made use of the shared facilities of the Materials Research Science and Engineering Center (MRSEC) at UC Santa Barbara: NSF DMR–2308708. The UC Santa Barbara MRSEC is a member of the Materials Research Facilities Network (www.mrfn.org).

\section{Author Contributions}

R.S., S.D., and R.S.R. conducted the optical measurements and analyzed the data. K.Y. and J.G.A. synthesized and characterized the samples. J.W.H. conceived and designed the project. R.S. and J.W.H. wrote the manuscript, with input from all authors.

\section{Competing Interests}

The authors declare no competing interests.

\end{document}